\documentclass[prx,twocolumn,preprintnumbers,amsmath,amssymb,superscriptaddress]{revtex4} 
\usepackage[T1]{fontenc}
\usepackage[utf8]{inputenc}
\usepackage[german,english]{babel}
\usepackage{epsfig}
\usepackage{amssymb}
\usepackage{amsmath}
\usepackage{braket}
\usepackage{bm}
\usepackage{color}
\usepackage{times}
\usepackage[colorlinks,bookmarks=false,citecolor=blue,linkcolor=red,urlcolor=blue]{hyperref}
\usepackage{float}

\usepackage{soul}

\newcommand{\beq}{\begin{equation}}
\newcommand{\eeq}{\end{equation}}
\newcommand{\bea}{\begin{eqnarray}}
\newcommand{\eea}{\end{eqnarray}}

\newcommand\ag[1]{{\color{red}{\bf \small #1}}}

\begin{document}
\setstcolor{red}
\title{Prethermal quantum many-body Kapitza phases of periodically driven spin systems
}
\author{Alessio Lerose}
\affiliation{SISSA -- International School for Advanced Studies, via Bonomea 265, 34136 Trieste, Italy.}
\affiliation{INFN, Sezione di Trieste, via Bonomea 265, 34136 Trieste, Italy.}
\author{Jamir Marino} 
\email{jamirmarino@fas.harvard.edu}
\affiliation{Institut f\"{u}r Theoretische Physik, Universit\"{a}t zu K\"{o}ln, D-50937 Cologne, Germany\\
Department of Physics and Center for Theory of Quantum Matter, University of Colorado Boulder, Boulder, Colorado 80309, USA\\
Department of Physics, Harvard University, Cambridge MA 02138, United States}
\author{Andrea Gambassi}
\affiliation{SISSA -- International School for Advanced Studies, via Bonomea 265, 34136 Trieste, Italy.}
\affiliation{INFN, Sezione di Trieste, via Bonomea 265, 34136 Trieste, Italy.}
\author{Alessandro Silva}
\affiliation{SISSA -- International School for Advanced Studies, via Bonomea 265, 34136 Trieste, Italy.}



\begin{abstract}
 As realized by Kapitza long ago, a rigid pendulum can be stabilized upside down by periodically driving its suspension point with tuned amplitude and frequency. While this dynamical stabilization is feasible in a variety of instances in systems with few degrees of freedom, it is natural to search for generalizations to multi-particle systems. In particular, a fundamental question is whether, by periodically driving a single parameter in a many-body system, one can stabilize an
otherwise unstable phase of matter against all possible fluctuations of its microscopic degrees of freedom.
In this work we show that such stabilization occurs in experimentally realizable quantum many-body systems:
a periodic modulation of a transverse magnetic field can make ferromagnetic spin systems with long-range
interactions stably trapped around unstable paramagnetic configurations as well as in other unconventional
dynamical phases with no equilibrium counterparts. 
We demonstrate that these
quantum Kapitza phases have a long lifetime and can be observed in current experiments 
with trapped ions.
\end{abstract}

\date{\today}
\maketitle

\section{Introduction}

Periodic drivings are ubiquitous in natural phenomena and particularly in applications, ranging from electronics to condensed matter physics \cite{condmat, BukovReview, opticallatt}.
Understanding driven systems is of paramount importance
in the context of quantum technologies, since these systems can both 
realize peculiar phases of matter and help manipulating quantum information \cite{quantumtech}.
In fact, time-periodic protocols have been theoretically proposed and experimentally realized to engineer a variety of systems, including topological phases \cite{floquettopo}, time crystals \cite{timecrystalth,timecrystalexp}, exotic Bose-Einstein condensates \cite{exoticBEC}. 
All of them have no equilibrium counterparts, i.e., they do not exist in the absence of 
driving. For instance, a gas of bosons may condense in a non-uniform, $\mathbf{\pi}$-quasimomentum state in the presence of a rapidly varying electric field or of a shaken lattice \cite{exoticBEC}. 
Similarly, while invariance under time-translations cannot be broken at equilibrium, the formation of discrete time crystals under the effect of AC-driving has been theoretically proposed \cite{timecrystalth} and experimentally observed \cite{timecrystalexp}. 

In systems with few degrees of freedom a periodic drive might have spectacular effects, such as the stabilization of a pendulum upside down. A theory of this phenomenon was formulated by P. Kapitza in 1951 \cite{Kapitza}, which found applications in laboratories (see, e.g., 
Refs.~\cite{synchrotron,Paul}) as well as to the stabilization of otherwise unstable phases (referred to as \emph{Kapitza phases}) in many-body systems whose description, at the mean-field level, can be reduced to a few collective degrees of freedom  (see, e.g., 
Refs.~\cite{stabilizedBEC,spinmixing,Kapitzaexp,JosephsonKapitza}).
In this work, we demonstrate that this  
stabilization
may occur over a parametrically large time scale in Floquet prethermal phases of periodically driven quantum magnets affected by many-body fluctuations.

Although the results are general, we focus on experimentally relevant models, i.e.,
spin chains with long-range ferromagnetic interactions described by the Hamiltonian
\beq
\label{eq:generalIsing}
H= - \sum_{i\ne j}^N \frac{J}{\lvert i - j \rvert^{\alpha}}\sigma^x_{i} \sigma^x_{j} - B \sum_{i}^N \sigma^z_{i},
\eeq

\begin{figure}[h!]
 \includegraphics[width=0.48\textwidth]{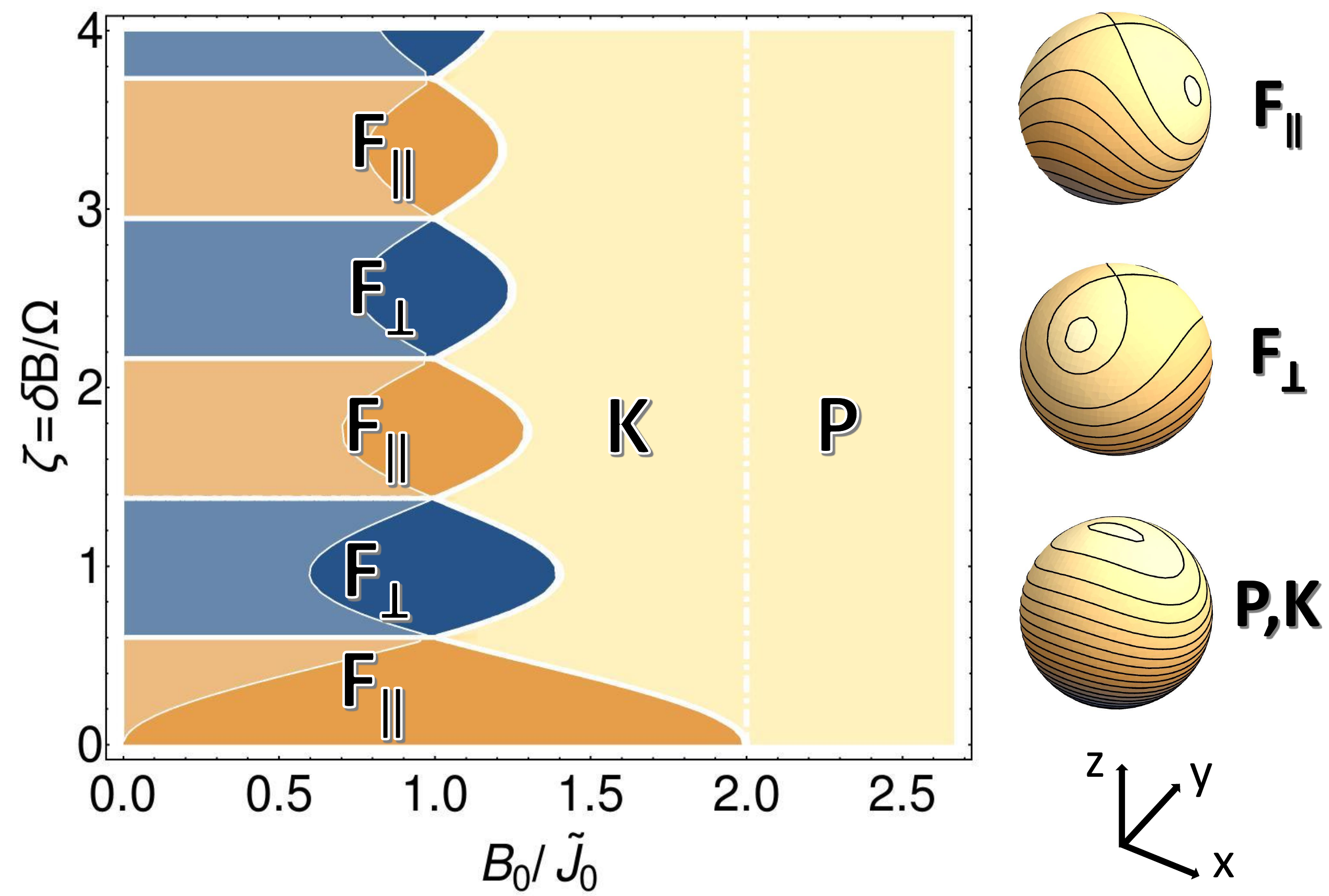} 
\caption{Left: Fast-driving non-equilibrium phase diagram of the periodically driven long-range Ising model defined by Eqs.~\eqref{eq:generalIsing} and \eqref{eq:protocol}. Upon varying the average magnetic field $B_0$ and the rescaled modulation amplitude $\zeta = \delta B / \Omega$, a dynamical paramagnetic phase $P$, a dynamically stabilized Kapitza paramagnetic phase  $K$, a conventional dynamical ferromagnetic phase $F_{\parallel}$ and an unconventional dynamical ferromagnetic phase $F_{\perp}$ with orthogonal magnetization emerge.
The axis $\zeta=0$ corresponds to the equilibrium phase diagram, where a ferromagnetic $F_\|$ and a paramagnetic $P$ phase are present.
The diagram shows the exact phase boundaries of the infinite-range system with $\alpha=0$. (Note that the dashed line separating $K$ and $P$ does not correspond to 
an actual phase transition.)
When $0<\alpha\le2$, quantum fluctuations modify
these boundaries, leaving however their qualitative structure unaltered.
Within the shaded region on the left, a second Kapitza phase coexists with $F_{\parallel,\perp}$, but is stable for $\alpha=0$ only.
Right: Schematic phase portraits on the Bloch sphere of the effective high-frequency Hamiltonians governing the evolution of the collective spin of the system, highlighting the various phases.
These phases persist up to time scales $\tau \sim \exp ( \, \text{const} \times \Omega/\tilde{J}_0)$ for finite driving frequencies $\Omega$ larger than the characteristic energy scale $\tilde{J_0}=\sum_{r}^N J/r^{\alpha}$ of the system, before eventual heating takes place.}
\label{fig:2}
\end{figure}

\noindent where $\sigma^{\mu}_{i}$'s are Pauli matrices, $\alpha$, $J > 0$ (when $0\le\alpha<1$ the scaling $J\propto N^{\alpha - 1}$ yields a meaningful thermodynamic limit \cite{reviewLR}), and the magnetic field is periodically varied,
\beq 
\label{eq:protocol} B(t)=B_0 + \delta B \, \cos(\Omega t).
\eeq
These systems 
accurately model the non-equilibrium dynamics
of quantum simulators with trapped ions ($0 < \alpha < 3$) \cite{Monroe,Blatt} or Rydberg atoms ($\alpha = 6$) \cite{Lukin,Browaeys}. 
In Fig.~\ref{fig:2}  our main findings are illustrated: The non-equilibrium phase diagram of the driven systems features a number of dynamically stabilized phases, which include many-body analogs of the Kapitza pendulum as well as unconventional magnetically ordered phases with no equilibrium counterpart.

\smallskip

 \section{Infinite-range systems}
 \label{sec:infiniterange}
 
\begin{figure}[t]
 \includegraphics[width=.48\textwidth]{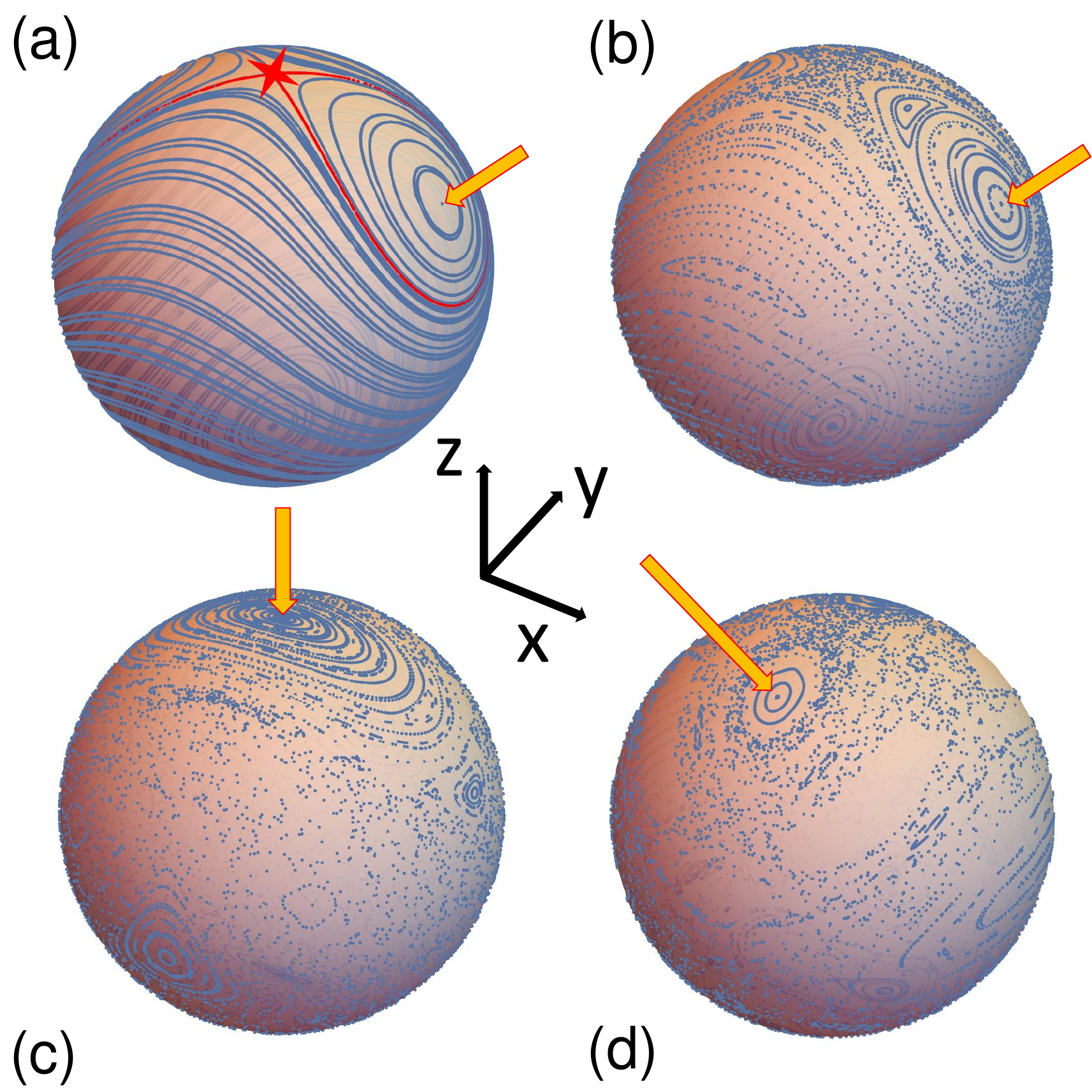} 
\caption{%
  Dynamics on the Bloch sphere of the infinite-range ($\alpha=0$) ferromagnet in the thermodynamic limit.
  (a) Semiclassical phase space trajectories of the static Hamiltonian with $B/\tilde{J}_0=1.2$. 
  (b), (c), (d): Stroboscopic trajectories $\{\vec{S}(t_n)\}$, with $t_n=2\pi n/\Omega$, $n=0,1,2,\dots$ of the semiclassical collective spin on the Bloch sphere with driving frequency $\Omega/\tilde{J}_0=5$ and increasing $\delta B/\tilde{J}_0=0.01$ (b), 3.3 (c), and 5 (d), with $B_0/\tilde{J}_0=1.2$. 
Panel (b) shows the presence of a possible ferromagnetic dynamical ordering, corresponding to the evolution occurring within a single ferromagnetic sector $S_x>0$, with a special synchronized trajectory (appearing as a single point under stroboscopic observations), together with the onset of chaotic behavior around the unstable paramagnetic point \cite{Russomanno}. Panel (c) shows the appearance of a dynamically stabilized phase, akin to the well-known stabilization of the inverted driven Kapitza pendulum \cite{Kapitza,Landau1}. 
Panel (d) shows that for larger driving frequencies, an unconventional dynamical ferromagnetic ordering appears, where the direction of the magnetization is orthogonal to
the direction $x$ of the actual ferromagnetic interactions.
Islands with stable stroboscopic trajectories
are indicated by the arrows.
}
\label{fig:1}
\end{figure}
 
In order to gain insight into the physics of this problem it is worth to start by analyzing the simplest limit $\alpha\to0$, where mean-field theory becomes exact and ideas analogous to those employed by Kapitza in 1951 can be applied \cite{Kapitza}. 
In fact, in this limit $H$ in Eq.~\eqref{eq:generalIsing} reduces to the Lipkin-Meshkov-Glick (LMG) model, which is equivalent to a single macroscopic spin \cite{LMG} on the Bloch sphere.  Indeed, the coupling strength is the same $J = \tilde{J}_0/N$ for all pairs of spins, hence $H$ describes the dynamics of a single collective spin $\vec{S}=\sum_i \vec{\sigma}_i /N$ \cite{SciollaBiroliMF,Zunkovic,BrandesLMG}. 
In the thermodynamic limit $N\to\infty$ the rescaled Hamiltonian $H/N$ becomes equivalent to its classical limit
$\mathcal{H}_{\text{cl}} = - \tilde{J}_0 \, S_x^2 - B \, S_z$.
At zero temperature and constant $B$, this system has a paramagnetic phase for $\lvert B\rvert>2\tilde{J}_0$, where all the microscopic spins are oriented along the transverse direction $z$ of the field, and a ferromagnetic phase for $\lvert B\rvert<2\tilde{J}_0$, where the spins acquire a non-vanishing component along the longitudinal direction $x$ \cite{SciollaBiroliMF,DasSengupta,Zunkovic}.

\subsection{Dynamics}

 The non-equilibrium evolution of the system in the presence of a time-dependent field 
$B = B (t)$
 is described by the dynamics of the collective spin $\vec{S}(t)$ on the sphere of radius $1$ governed by the classical Hamiltonian $\mathcal{H}_{\text{cl}}(t)$. 
When $B$ is static and supports the ferromagnetic state indicated by the arrow in Fig.~\ref{fig:1}(a), $\vec{S}(t)$ follows one of the trajectories represented on the Bloch sphere in panel (a), selected by the initial condition $\vec{S}(0)$ (for definiteness, we will assume $B_0\ge 0$ throughout). 
Two families of them are characterized by a ferromagnetic-like, symmetry-breaking periodic evolution with opposite signs of the non-vanishing time-averaged order parameter $\overline{S_x}$. A trajectory (red) passing through the equilibrium unstable paramagnetic point (red star) separates these two families from the paramagnetic-like orbits with $\overline{S_x}=0$.
Turning on the modulation as in Eq. \eqref{eq:protocol}, representative samples of discrete stroboscopic trajectories $\{\vec{S}(t_n)\}$, where $t_n=2\pi n/\Omega$, with $n=0,1,2,\dots$ of the semiclassical collective spin are reported in Fig.~\ref{fig:1}(b), (c), and (d).
When the modulation $\delta B$ is small [see panel (b)], the ferromagnetic 
states leave room to periodic trajectories of the collective spin within the corresponding ferromagnetic sector \emph{synchronized} with the drive (and hence appearing as a single point under stroboscopic observations). 
 Conversely, initial states in a neighborhood of the unstable paramagnetic point [red star in panel (a)] display chaotic motion as soon as $\delta B \neq 0$  \cite{Russomanno,DasSengupta}. 
As $\delta B$ increases, this chaotic region invades an increasingly large portion of the sphere \cite{Russomanno}. 
This behavior can be understood on the basis of classical KAM and chaos theory \cite{chaos,KAM} and related phenomena have been experimentally observed with Bose-Einstein condensates \cite{PoincareBirkhoff}.
(The structure of resonances in driven collective systems has been studied also in Refs. \cite{BrandesLMG,BrandesDicke}.)
Upon further increasing the modulation [see panel (c)], a region in the parameter space emerges where \emph{dynamical stabilization} of the unstable paramagnetic point occurs, thereby opening up a stability region around it. This phenomenon is analogous to the stabilization of the inverted pendulum discovered by Kapitza \cite{Kapitza,Landau1}. 
In addition to this Kapitza-like stabilization, as $\delta B$ increases with $B_0 \approx \tilde{J}_0$ 
[see panel (d)], an unconventional regime appears characterized by dynamical ferromagnetic ordering in the $yz$-plane, orthogonal to the direction $x$ of the actual ferromagnetic interactions. 

\subsection{Fast-driving limit}  

In order to understand later on the full many-body case ($\alpha\ne0$), we 
first analyze
the behavior of the system described above in the regime of fast-driving $\Omega\to\infty$ as a function of the rescaled amplitude $ \zeta = \delta B / \Omega$. In fact, the effective Floquet 
Hamiltonian governing the stroboscopic evolution \cite{SpagnoliReview} can be determined non-perturbatively, by switching to a convenient oscillating reference frame \cite{BukovReview} (see Appendix \ref{app:Magnus}). The effect of the driving then amounts to redistributing the ferromagnetic coupling strength along the  directions $x$ and $y$, thereby  turning the Ising model into an XY model, with
\beq
\label{eq:effXY_0}
 \mathcal{H}_{\text{eff}} = - \tilde{J}_0\left(\frac{1+\gamma(\zeta)}{2}  S_x^2 + \frac{1-\gamma(\zeta)}{2}  S_y^2\right) - B_0 \, S_z,
 \eeq
and anisotropy parameter $\gamma(\zeta)={\cal J}_0(4\zeta)$, where ${\cal J}_0$ is the Bessel function of the first kind.

As $\zeta$ increases from zero, the effective ferromagnetic interaction along $x$ weakens, which makes it possible to dynamically stabilize the paramagnetic configuration. 
The exact boundary $B_0 = B_{\text{cr}}(\zeta) \equiv  \tilde{J}_0 (1 + \lvert {\cal J}_0(4\zeta) \rvert)$ of the Kapitza phase $K$ is reported in Fig.~\ref{fig:2}. 
Note that this region is continuously connected  with the paramagnetic one \ag{$P$} in the phase diagram, see Fig.~\ref{fig:2}, similarly to the region of dynamical stabilization of the Kapitza pendulum, which is continuously connected with the parameter region with a reversed direction of gravity, in which stability is trivial \cite{Landau1}.
%
As $\zeta$ increases even further, due to the oscillations of ${\cal J}_0$ around zero, intervals with a negative anisotropy $\gamma$ appear, making ferromagnetic ordering along the direction $y$ become favored.
The mechanism is thus elucidated for the occurrence of the
unconventional dynamical phases with ferromagnetic ordering in the $yz$-plane, orthogonal to the direction $x$ of the actual ferromagnetic interaction, which builds up whenever $\gamma<0$, $B_0<\tilde{J}_0(1-\gamma)$, i.e., within
the regions denoted by $F_{\perp}$ in Fig.~\ref{fig:2}.
A second Kapitza phase coexists with $F_{\parallel,\perp}$ for $B_0 <\tilde{J}_0(1-\lvert {\cal J}_0(4\zeta) \rvert)$, as discussed in Appendix \ref{app:CoexistenceKF}.

The numerical results reported in Fig.~\ref{fig:1} show that these non-equilibrium phases persist even at smaller driving frequencies, comparable to the characteristic energy scale $\tilde{J}_0$ of the system. Indeed, as discussed in Appendix \ref{app:Magnus}, when the driving frequency $\Omega$ is large but finite,  the effective Floquet Hamiltonian \eqref{eq:effXY_0} receives perturbative corrections in an expansion in inverse powers of $\Omega$. The first term beyond Eq. \eqref{eq:effXY_0} is reported in Appendix \ref{app:FiniteFreq} [cf. Eq. \eqref{eq:effXY_1}] and it causes small quantitative modifications of the boundaries in Fig. \ref{fig:2}.
Furthermore, this dynamical stabilization is robust to finite-size effects, as demonstrated in Sec.~\ref{sec:ed} below.

\smallskip

\section{Variable-range interactions} 
The behavior of infinite-range systems can essentially be understood in terms of one-body physics. However, when interactions  have a non-trivial spatial 
dependence, fluctuations at all length scales are activated. 
The possibility to stabilize many-body dynamical phases by modulating in time a global external field represents a major conceptual and practical challenge.
In order to address this problem, we study below the spin system \eqref{eq:generalIsing} with $\alpha\neq 0$ and we will show how the dynamical phases reported above can be stabilized also for $0<\alpha\le2$, where quantum fluctuations around the semiclassical evolution are not suppressed, with the exception of the coexistence region.
Their effect is reduced by decreasing the parameter $\alpha$,
which  continuously connects the models with their infinite-range semiclassical limit.

When $\alpha \ne 0$, both the total spin $\vec{S}$, corresponding to the $k=0$ Fourier mode of $\vec{\sigma}_i$, and all the $k\ne0$ quasi-particle (\textit{spin wave}) excitations are affected by interactions \cite{LeroseShort,LeroseLong}.
In order to account for the coupled dynamics of the collective spin and of the spin wave excitations around the time-dependent direction of $\vec{S}(t)$,
we employ the \emph{time-dependent spin wave theory} developed in Ref. \cite{LeroseShort}, see also Appendix \ref{app:TDSWT} for a concise overview.
In the presence of $k\ne0 $ modes, representing all microscopic fluctuations, the system may be thought of as a macroscopic semiclassical collective degree of freedom, i.e., the total spin $\vec{S}(t)$, which ``drags'' along an extensive set of quantum oscillators $(\tilde{q}_k,\tilde{p}_k)$'s, i.e., of microscopic degrees of freedom corresponding to the bosonic spin wave excitations with quasi-momentum $k\ne0$ \cite{LeroseShort,LeroseLong}. Indeed, the time-dependent spin wave theory maps spin fluctuations into such bosonic excitations and the Hamiltonian $H(t)$ is then written
 in terms of the collective spin variables $\vec{S}/\lvert\vec{S}\rvert=(\sin\theta \cos\phi, \sin\theta\sin\phi,\cos\theta)$ and of the spin wave operators $\tilde{q}_k$'s, $\tilde{p}_k$'s. Truncation to quadratic order in the quantum fluctuations yields
\begin{multline}
\label{eq:timeindepH}
H(t) =  - N B(t) (1-\epsilon)\cos\theta  - N \tilde{J}_0 \big[  (1-\epsilon) \sin\theta \cos \phi \big]^2  \\
        -  4 \sum_{k\ne0}\tilde{J}_k \bigg(
       \cos^2\theta \cos^2 \phi \; \frac{\tilde{q}_k \tilde{q}_{-k}}{2} + \sin^2 \phi \; \frac{\tilde{p}_k \tilde{p}_{-k}}{2} \\
       -  \cos\theta \cos \phi \sin \phi \; \frac{\tilde{q}_k \tilde{p}_{-k} + \tilde{p}_k \tilde{q}_{-k}}{2}
       \bigg),
\end{multline}
where $\tilde{J}_k$ is the Fourier transform of the interaction $J/r^\alpha$ 
(the scaling of $J$ as $N\to\infty$, see below Eq. \eqref{eq:generalIsing}, guarantees that $\tilde{J}_0$ is finite in the thermodynamic limit)
and $\epsilon=  
\sum_k (\tilde{q}_k \tilde{q}_{-k} + \tilde{p}_k \tilde{p}_{-k}-1)/N $ is the relative depletion of the total spin length from its maximal value, 
i.e., $\lvert \vec{S} \rvert=1-\epsilon$. The last term in Eq.~\eqref{eq:timeindepH} accounts for the interaction between the collective semiclassical spin $\vec{S}$ and the quantum spin wave excitations.

\subsection{ Dynamically stabilized many-body  phases} 

\begin{figure}[t]
\centering

\includegraphics[width=0.46\textwidth]{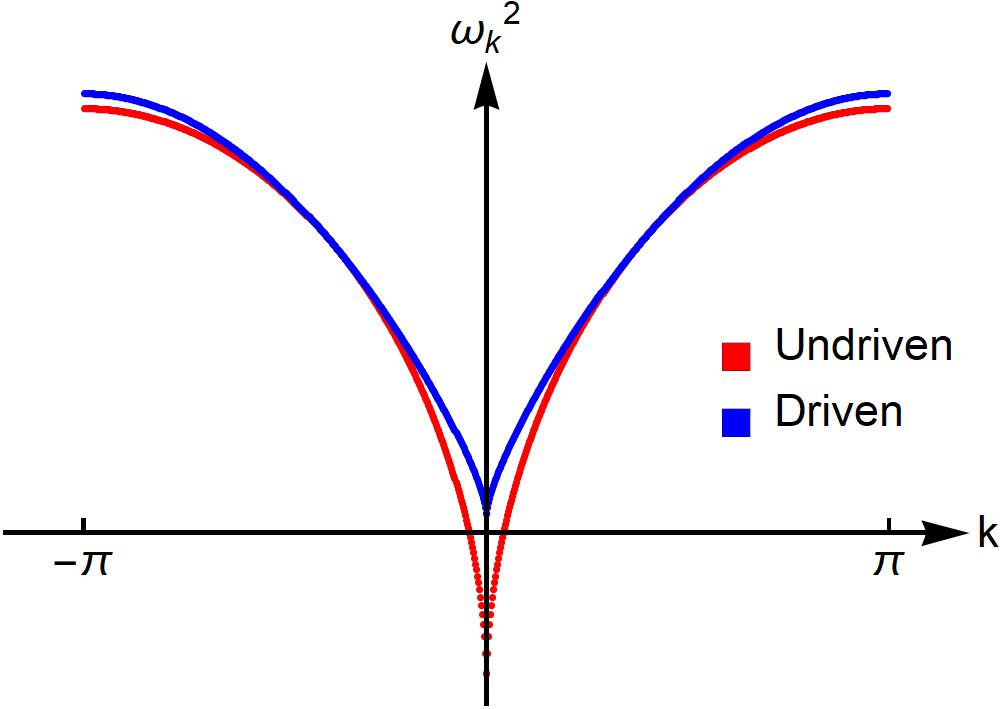} 
\caption{%
Stabilization of the \emph{many-body} Kapitza phases. In the presence of a suitable periodic driving, the otherwise unstable spectrum of quantum excitations 
around the paramagnetic configuration gets simultaneously dynamically stabilized for all values of $k$.
Here $\alpha=1.5$, $N=400$, and $B_0/\tilde{J}_0=1.35$: in the absence of the driving $\delta B = 0$, the system is in the ferromagnetic phase. 
The red points represent the (squared) frequency spectrum $\omega_k^2 = 4 B_0(B_0-2\tilde{J}_k)$ of the spin wave excitations, labeled by their wavevector $k$, with an extended interval of \emph{unstable} long-wavelength modes (i.e., $\omega_k^2 < 0$ for $k$ near $0$). 
As the driving is turned on with a strength in a suitable range of values, not only the collective spin mode with $k=0$ discussed in Sec. \ref{sec:infiniterange}, but also the whole set of modes with $k\ne0$ become \emph{stable} (i.e., $\omega_k^2 > 0$ for all $k$). 
The blue points show the exact effective dispersion relation $\omega_k^2 = 4 (B_0-\tilde{J}_k)^2$ in the presence of a high-frequency driving $\Omega\to\infty$ with $\zeta=\delta B/\Omega=0.6014$ (corresponding to $\gamma=0$ in the effective Hamiltonian, see the text). 
When $\tilde{J}_0 \ll \Omega < \infty$, this effective dispersion relation receives perturbative corrections in inverse powers of $\Omega$, thus no qualitative changes occur as long as the system is in the pre-thermal regime (see Sec. \ref{sec:heating} and references therein). 
As discussed in the text, upon decreasing $\alpha$ below $1$, the Fourier transform $\tilde{J}_{k}$ of the ferromagnetic couplings approaches zero for all $k\ne0$ in the thermodynamic limit, thus reducing this many-body problem to the single-body dynamical stabilization of the collective spin discussed in Sec. \ref{sec:infiniterange}.}
\label{fig:manybodystabilization}
\end{figure}

A many-body Kapitza phase consists of a simultaneous dynamical stabilization of the whole spectrum of quantum excitations around an unstable paramagnetic configuration. Intuition on this phenomenon can be obtained at the level of linear stability by expanding $H(t)$ to quadratic order in the quantum fluctuations, as in Eq.~\eqref{eq:timeindepH}, around the point $\theta=0$:
\beq
\label{eq:parpoint}
H(t) = E(t) +
        2 \sum_{k}  \left[ ( B(t)-2\tilde{J}_k )
        \frac{\tilde{q}_k \tilde{q}_{-k}}{2} + B(t) \frac{\tilde{p}_k \tilde{p}_{-k}}{2}
       \right],
\eeq
where $E(t)=-2N B(t)$ and $k=2\pi n / N$ with $n=0,1,\dots,N-1$ (assuming periodic boundary conditions for simplicity). In the absence of modulation in the ferromagnetic phase  [i.e., $B(t)=B_0<2\tilde{J}_{0}$], an extended interval near $k=0$ in the spin waves band  corresponds to unstable modes,
as their corresponding frequency $\omega_k = 2[B_0(B_0-2\tilde{J}_k)]^{1/2}$ becomes imaginary for $\tilde{J}_k>B_0/2$. However, upon introducing the modulation $B(t)$ as before, the effective dispersion relation  is modified, and for a suitable choice of the driving parameters $\omega_k$ may become real for all values of $k$. The occurrence of this non-trivial stabilization of an otherwise unstable phase of matter against all possible fluctuations of its degrees of freedom is illustrated in Fig. \ref{fig:manybodystabilization} and it represents an actual generalization of the Kapitza pendulum to a genuine many-body system.
%
\begin{figure}[h!]
 \includegraphics[width=0.45\textwidth]{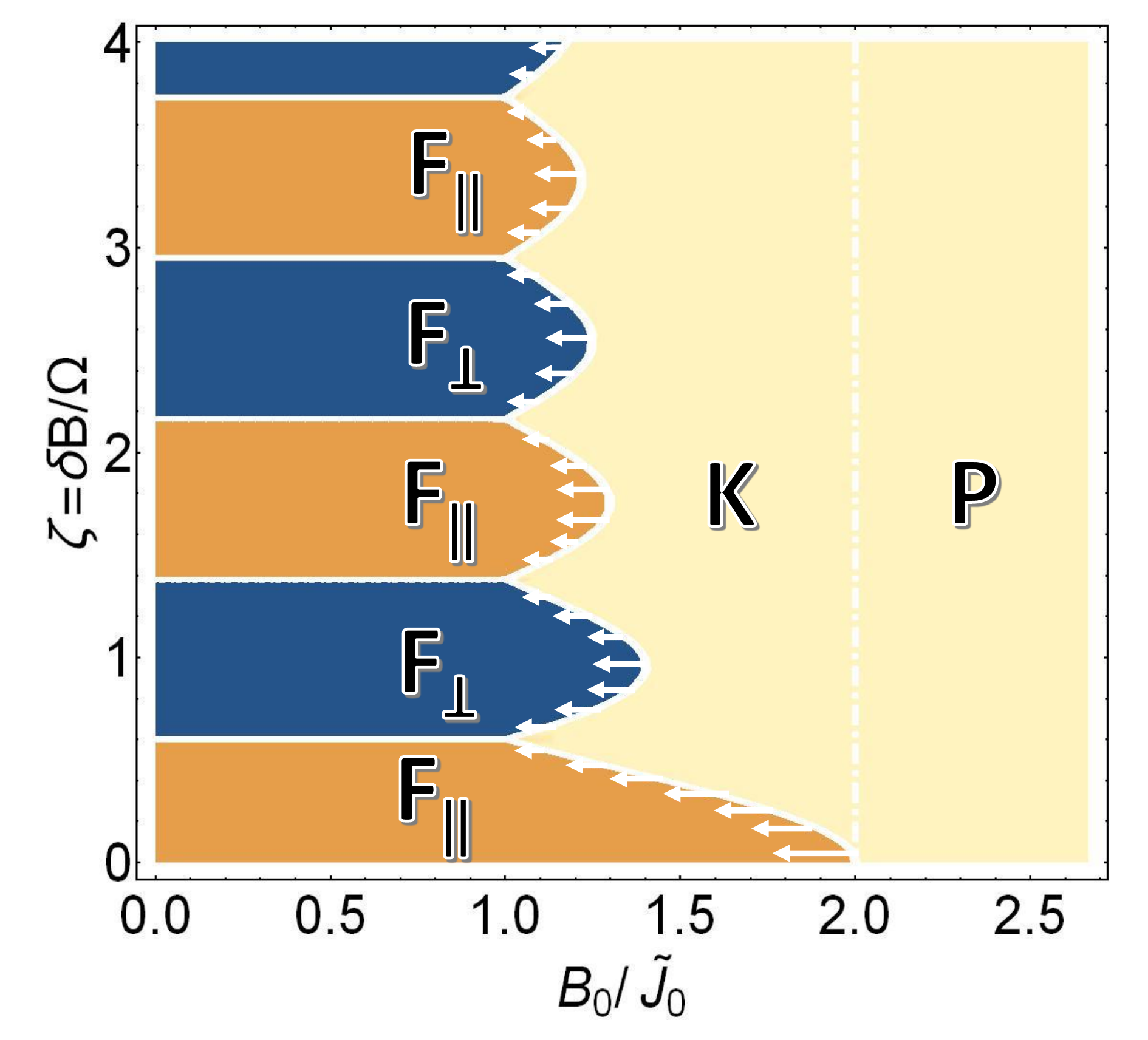} 
\caption{Fast-driving non-equilibrium phase diagram of the periodically driven long-range Ising model defined by Eqs.~\eqref{eq:generalIsing} and \eqref{eq:protocol}, for $\alpha>0$.
Compared to Fig. \ref{fig:2}, the shaded region of coexistence of phase $K$ with $F_{\parallel,\perp}$ has disappeared, and the left boundary of region $K$ moves leftwards upon increasing $\alpha$, as determined by Eq. \eqref{eq:shift} and indicated by the white arrows. This displacement is vanishingly small in the thermodynamic limit for $0<\alpha\le1$, due to the rescaling of long-range interactions as $N\to\infty$ [see below Eq. \eqref{eq:generalIsing}], whereas it is finite for $\alpha>1$. The amount indicated by the arrows corresponds to Eq. \eqref{eq:shift} with $\alpha=1.5$ (it is magnified by a factor of $2$ for ease of visualization).}
\label{fig:NEQphasediagramAlpha}
\end{figure}
%

\begin{figure*}[t]
\centering
\includegraphics[width=\textwidth]{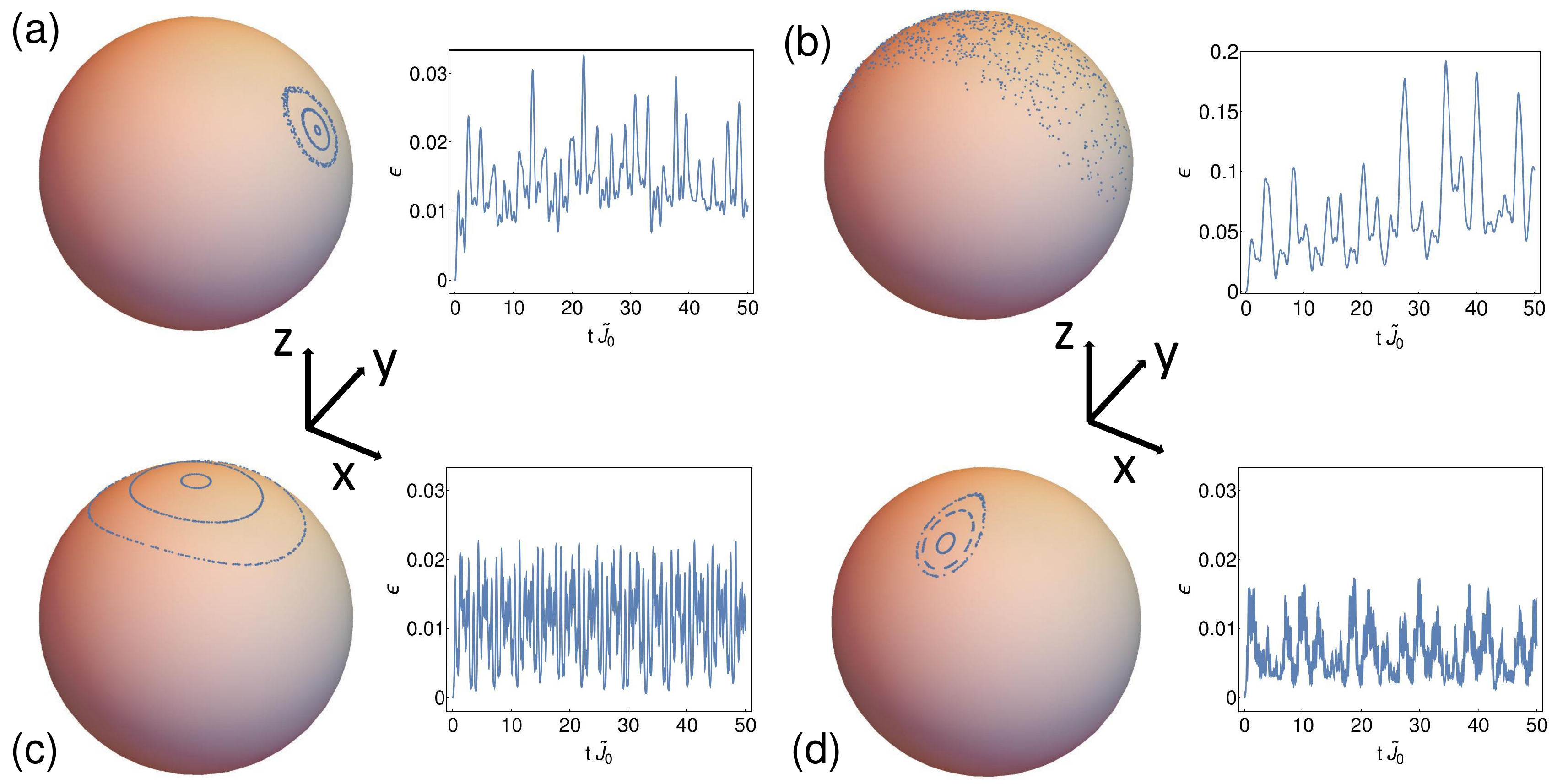} 
\caption{%
Persistence of the dynamically stabilized phases with finite driving frequency.
Left in each panel: Stroboscopic time-evolution of the total spin (projected on the unit sphere) of the long-range Ising chains in 
Eq.~\eqref{eq:generalIsing} with $\alpha\ne0$, subject to the modulated magnetic field in Eq.~\eqref{eq:protocol}. 
The dynamics is obtained by numerically integrating the system of coupled evolution equations for the total spin and the spin waves provided by the time-dependent spin wave theory, see Eq.~\eqref{eq:timeindepH} and Appendix \ref{app:TDSWT}.
In all simulations, the static field is $B_0/\tilde{J}_0=1.2$, as in Fig.~\ref{fig:1}, the driving frequency is $\Omega/\tilde{J}_0=8$, the system size is $N=100$, and the system is initialized in spin-coherent (fully polarized) states in the $xz$ (panels (a), (b), (c)) and $yz$ (panel (d)) planes. 
Right in each panel: relative departure $\epsilon(t)$ of the total spin from its maximal length $N/2$ [i.e., $  \lvert \vec{S} (t) \rvert = 1 - \epsilon(t)  $], due to the generation of quantum spin wave excitations, corresponding to the largest trajectory in each panel. Notice $\epsilon(t=0) =0$ with our choice of fully polarized initial states. 
In particular: 
(a) Dynamical ferromagnetic phase, with 
$\alpha=1$ and $\delta B/\tilde{J}_0 = 0.05$. 
(b) Fast heating in the chaotic dynamical regime, with   
 $\alpha=0.8$, $\delta B/\tilde{J}_0 = 0.2$. 
(c) Dynamically stabilized Kapitza phase, with 
 $\alpha=1$, $\delta B/\tilde{J}_0 = 5.33$. 
(d) Unconventional, dynamically stabilized ferromagnetic  phase with magnetization in the $yz$-plane orthogonal to the direction $x$ of the actual ferromagnetic interactions, with $\alpha=1$, $\delta B/\tilde{J}_0 = 8$. 
Panels (a), (c), and (d) demonstrate that the dynamical phases $F_{\parallel}$, $K$, $F_{\perp}$ (see Fig.~\ref{fig:2}), respectively, continue to exist at finite driving frequency. 
The amount of excitations generated remains small and the total energy remains bounded across many cycles, qualifying these phases as being \emph{prethermal}.
In panel (b), instead, the broad frequency spectrum of the chaotic semiclassical motion gives rise to resonant generation of excitations, witnessed by the growth of $\epsilon(t)$ (notice the different vertical scale in the plot), and absorption of energy from the drive (\emph{heating}). 
The heating rate in this case increases upon increasing $\alpha$.}
\label{fig:3}
\end{figure*}

In order to understand how all the degrees of freedom can get  dynamically and simultaneously stabilized by driving a single global field $B(t)$,
we concentrate first on the fast-driving limit $\Omega\to\infty$ as a function of the rescaled driving amplitude $\zeta$, which can be studied analytically also for $\alpha\neq 0$.
Here, the effective Floquet  Hamiltonian governing the stroboscopic time-evolution is the long-range XY spin chain,
\beq
\begin{split}
\label{eq:effXY_alpha}
H_{\text{eff}} 
=&
 -  \sum_{i\ne j}^N \frac{J}{\lvert i-j \rvert^{\alpha}} 
 \bigg(
\frac{1+\gamma(\zeta)}{2} \sigma^x_{i} \sigma^x_{j} + \frac{1-\gamma(\zeta)}{2} \sigma^y_{i} \sigma^y_{j} 
 \bigg)  \\
 & \qquad \qquad \qquad \qquad \qquad\qquad\qquad\;
 - B_0 \sum_{i}^N \sigma^z_{i},
 \end{split}
\eeq
where the parameter $\gamma(\zeta)$ is the same as in Eq. \eqref{eq:effXY_0} and is independent of the particular dependence of the interactions on the distance (see Appendix \ref{app:Magnus}).

The stability analysis of the paramagnetic configurations is carried out by expanding $H_{\text{eff}}$ at the quadratic order in the spin wave operators around the field direction $z$ and hence by determining the range of parameter values within which the dispersion relation is real. It turns out that
a simultaneous dynamical stabilization of the whole spectrum of spin wave excitations, such as that illustrated in Fig. \ref{fig:manybodystabilization}, is possible within the region denoted by $K$ in Fig. \ref{fig:2}, and, upon increasing $\alpha$, the quantum fluctuations solely modify its phase boundary. Within the shaded region in Fig. \ref{fig:2}, instead, the second Kapitza phase turns out to be unstable to many-body fluctuations at finite wavelength: although the driving stabilizes the collective mode with $k=0$,  an extended interval in the Brillouin zone with $k\ne0$ appears, which is characterized by imaginary frequencies (see Appendix \ref{app:CoexistenceKF} for further details).
To the  lowest order in $\tilde{J}_{k\ne0}$, the shift $\Delta B_{\text{cr}}(\zeta)$ of the left boundary $B_0=B_{\text{cr}}(\zeta)$ of region $K$ is  given by 
\beq
\label{eq:shift}
\Delta B_{\text{cr}}(\zeta) = - \tilde{J}_0 \lvert \gamma(\zeta) \rvert \Bigg[ \int_{-\pi}^\pi  \frac{dk}{2\pi} \, \bigg(\frac{\tilde{J}_k}{\tilde{J}_0}\bigg)^2 \Bigg] \frac{1+\frac{3}{2} \lvert\gamma(\zeta)\rvert}{1+\lvert\gamma(\zeta)\rvert} ,
\eeq
as derived in Appendix \ref{app:shift}.
Due to the rescaling of long-range interactions  in the thermodynamic limit [see below Eq. \eqref{eq:generalIsing}], one has $\tilde{J}_{k\ne0} \to 0$ when $0<\alpha\le1$ (i.e., as $N\to\infty$ fluctuations are suppressed and the system becomes equivalent to its infinite-range limit), whereas $\tilde{J}_{k\ne0}$  approaches a finite value when $1<\alpha\le2$, with a cusp behavior $\tilde{J}_{k\ne0} \thicksim \tilde{J}_{0}(1-c \lvert k \rvert^{\alpha-1})$ for small wavenumbers $k$.
Therefore, the modification of the phase boundary in Fig. \ref{fig:2} due to quantum fluctuations is vanishingly small in the thermodynamic limit when $0<\alpha\le1$ and 
is finite as $\alpha> 1$, the smallness parameter being $\alpha-1$.
The resulting modified non-equilibrium phase diagram is presented in Fig. \ref{fig:NEQphasediagramAlpha}.

This proves the \emph{existence} of many-body non-equilibrium Kapitza phases for 
sufficiently fast driving (and in the regime of prethermal slow heating, see 
Sec.~\ref{sec:heating} below), under the sole condition that the effect of fluctuations is not so strong as to modify the bulk structure of the equilibrium phases of the effective Hamiltonian 
$H_{\text{eff}}$, which is known to be generally the case for long-range interactions with 
exponent $\alpha \le 2$, as well as for higher-dimensional systems with short-range interactions \cite{reviewLR,SFT}. 

More generally, the stability of the various many-body non-equilibrium phases can be determined by studying the local extrema of the mean-field energy landscape of  $H_{\text{eff}}$ and the corresponding spectra of excitations.  In particular, driving amplitudes $\zeta$ corresponding to negative anisotropy parameters $\gamma(\zeta)<0$ allow the appearance of dynamically stabilized unconventional ferromagnetic phases with orthogonal ferromagnetic ordering in the $yz$-plane whenever $ B_0 < \tilde{J}_{0}(1-\gamma) + \Delta B_{\text{cr}}$ [cf. Eq. \eqref{eq:shift}], which have no equilibrium counterpart in the Ising model. Such phases arise under the same conditions as the Kapitza phases discussed above.

\subsection{Prethermalization and heating} 
\label{sec:heating}

We address the footprint of the fast-driving non-equilibrium phase diagram on the finite-frequency dynamics, upon reducing $\Omega$ down to a scale comparable with the microscopic energy scale $\tilde{J}_{0}$ of the system.
In this case one should expect the system to eventually absorb an ever-increasing amount of energy from the drive \cite{DAlessioRigol,prethermhighfreq}. In order to address  this point, we initialize the system in various fully polarized states parameterized by angles $(\theta_0,\phi_0)$ on the Bloch sphere, and study the out-of-equilibrium evolution for various values of $\alpha>0$ and driving parameters $B_0,\delta B,\Omega$ by
 numerically integrating the dynamical equations of the time-dependent spin wave theory, where the heating rate can be monitored, e.g., through the depletion of the collective spin's magnitude from its maximal value (see Appendix \ref{app:TDSWT}). The results are illustrated in Fig.~\ref{fig:3}.

Whenever the system is initialized in a non-chaotic dynamical regime, $P$/$K$ or $F_{\parallel,\perp}$, and the frequency $\Omega$ is off-resonant with the spin-wave band, i.e., $\Omega \gg 4 \tilde{J}_{0}$, as shown in Fig.~\ref{fig:3}(a),(c),(d) the evolution presents a long time interval during which the absorption of energy from the drive, as well as the amount of spin-wave excitations, is bounded.
On the other hand, whenever the system is in a chaotic dynamical regime as in Fig.~\ref{fig:3}(b), irrespective of the value of $\Omega$ and of  $\alpha$, the amount of spin-wave excitations generated and the energy increase at a finite rate. Such a behavior corresponds to \emph{heating}, which has been extensively proven to be the generic response of a many-body system to an external periodic driving \cite{DAlessioRigol,WeidingerKnap,prethermhighfreq}, in the absence of dissipative mechanisms \cite{dissipative}. 
In the non-chaotic dynamical regimes $F_{\parallel,\perp}$ of panels (a) and (d), the synchronized trajectories of the collective spin $\vec{S}(t)$ act as an ``internal'' periodic driving at frequency $\Omega$ on the quantum oscillators $(\tilde{q}_k,\tilde{p}_k)$'s through the last interaction terms in the spin-wave Hamiltonian \eqref{eq:timeindepH}. As long as $\Omega$ is off-resonant (see above), the spin waves behave like a periodically driven system of quasi-free particles, which relaxes to a periodic quasi-stationary state described by a stroboscopic generalized Gibbs ensemble \cite{PDIsingchain,PDfree}. The presence of non-linear spin-wave interactions cause the latter \textit{prethermal} stage \cite{Canovi,Bukov,WeidingerKnap,Citro,DAlessioPolkovnikov} to be ultimately followed by slow heating, after a parametrically long time $\tau$ which scales as 
$\tau \sim \exp (\, \text{const} \times \Omega/\tilde{J}_{0})$ \cite{prethermhighfreq}.
On the contrary, the occurrence of chaotic motion of the collective spin $\vec{S}(t)$  translates, as in panel (b), into an irregular, noisy ``internal'' driving of the spin waves through the last terms in Eq. \eqref{eq:timeindepH}, possessing a broad frequency spectrum, whereby the unavoidable resonances with the spin waves band together with the local instability trigger the process of internal dissipation and hence a much faster heating.

\begin{figure}[t]
\centering
\includegraphics[width=.48\textwidth]{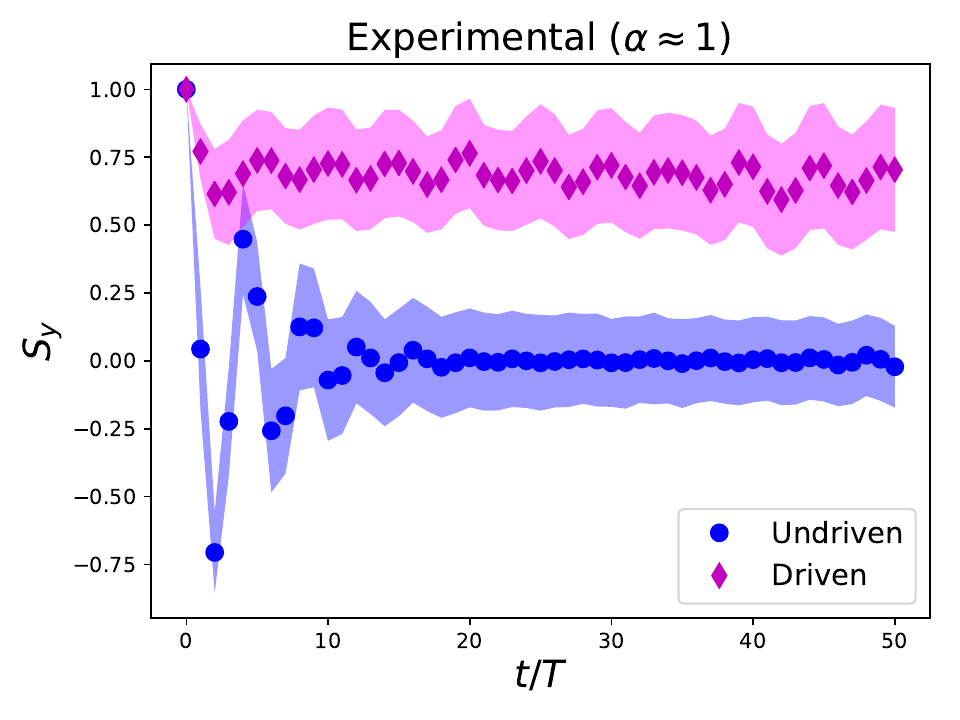} \\
\includegraphics[width=.48\textwidth]{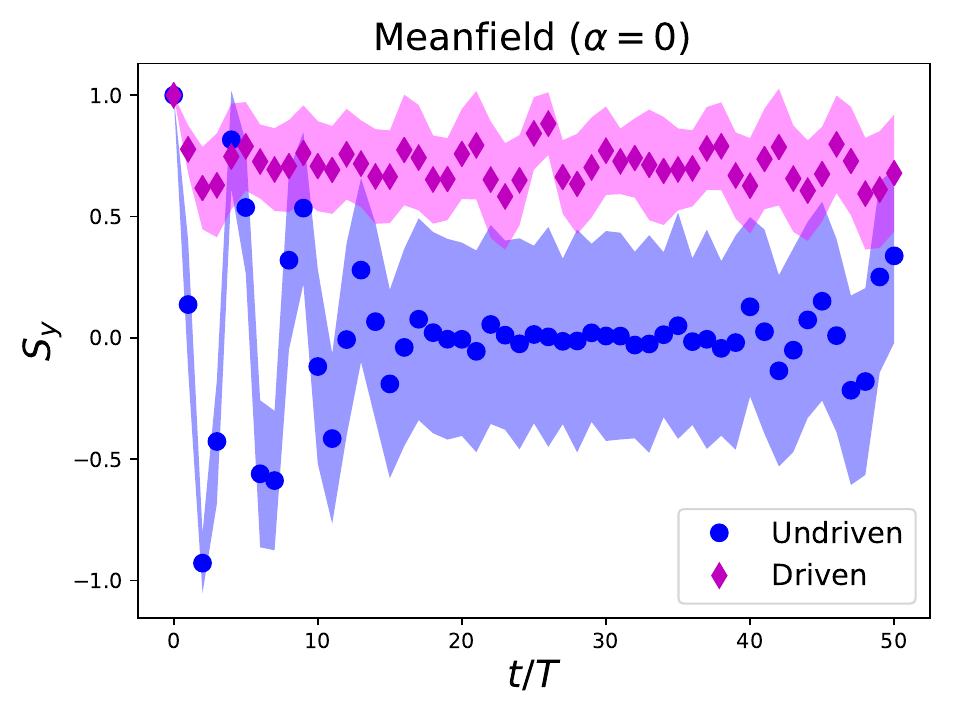}
\caption{%
Illustration of the dynamical stabilization of the unconventional ferromagnetic phase $F_{\perp}$ in a realistic trapped-ion setup (top panel) and in the corresponding infinite-range system (bottom panel). 
The orthogonal magnetization $\langle S_y(t) \rangle \pm \Delta S_y(t)$ (see the text) is plotted at stroboscopic times $t = nT$, with $T = 2\pi / \Omega$ and $n = 0,1,2,\dots$, for undriven and driven systems of $N=16$ spins 
with the actual space-dependent couplings $J_{ij}$ which characterize a chain of trapped ions experimentally studied in Ref.~\cite{Monroe}, roughly described by Eq.~\eqref{eq:generalIsing} with  $\alpha \approx 1$ and \eqref{eq:protocol} (top panel).
The system is initialized in a fully polarized state in the $y$-direction, and the driving parameters are $B_0/\tilde{J}_0=0.5$, $\zeta=1$, and $\Omega/\tilde{J}_0=8$, corresponding to a point well inside a region $F_{\perp}$ in Fig.~\ref{fig:NEQphasediagramAlpha}. 
For comparison, the corresponding simulation with $J_{ij}$ replaced by all-to-all uniform interactions of equal average strength is shown in the bottom panel. 
Ferromagnetic ordering in the $yz$-plane is dynamically stabilized by the drive, and is found to be robust to finite driving frequency, finite-size effects and ``imperfections''  in the long-range couplings, realizing a pre-thermal quantum many-body Kapitza phase observable with trapped-ion quantum simulators. 
}
\label{fig:experimental}
\end{figure}

\subsection{Quantum simulations with trapped ions.} 
\label{sec:ed}

We finally address the robustness of the dynamically stabilized non-equilibrium phases to finite-size effects as well as their observability in the setup of quantum simulations with trapped ions. 
In particular, we computed the non-equilibrium evolution of small driven systems of $N=16$ spins 
by numerically integrating the time-dependent many-body Schr\"{o}dinger equation, and demonstrate the occurrence of quantum many-body Kapitza phases for the long-range interacting chains in Eq. \eqref{eq:generalIsing} as well as with the space-dependent spin-spin couplings $J_{ij}$ which characterize 
a chain of trapped ions experimentally investigated in Ref.~\cite{Monroe}, roughly corresponding to model \eqref{eq:generalIsing} with $\alpha \approx 1$.

In Fig.~\ref{fig:experimental} we report an illustration of the stabilization of the \emph{ordered} phase $F_{\perp}$, which is of greater experimental relevance as it presents a type of magnetic ordering that is  absent in the equilibrium phase diagram. 
In order to probe its occurrence, the driving parameters in Eq. \eqref{eq:protocol} are chosen  well inside the region $F_\perp$ of the non-equilibrium phase diagram in Fig.~\ref{fig:NEQphasediagramAlpha}, and the system is initialized in a fully polarized state along the $y$-direction. 
In the top panel, the stroboscopic time-evolution of the orthogonal magnetization $\langle S_y(nT) \rangle$, with $T=2\pi/\Omega$ and $n=0,1,2,\dots$, 
is shown for the simulated trapped-ion system subject to the drive (magenta) as well as in the limit of vanishing driving strength (blue). In both cases, the shaded region around the symbols indicates the instantaneous quantum uncertainty of the magnetization $\Delta S_y(t) \equiv \sqrt{\langle S^2_y(nT) \rangle-\langle S_y(nT) \rangle^2}$. 
As the plot clearly shows, in agreement with the theory developed in this work, the drive stabilizes a magnetic ordering that is not possible in static conditions. 
We remark that the preparation of tilted fully polarized states, the implementation of the considered driving protocols and the measurements of the tilted magnetization can be achieved with standard experimental techniques, which makes it possible to actually observe this phenomenon with trapped ions.

In order to disentangle the finite-frequency and finite-size effects from the effects of having experimentally realistic interactions,
we report in the bottom panel of Fig.~\ref{fig:experimental} the outcome of analogous simulations in which the trapped-ion couplings have been replaced by uniform collective interactions $J_{ij}\equiv \tilde{J}_0/(N-1)$ with the same average strength. The qualitative appearance of the two plots is similar, in agreement with our theory. However, we observe that the strong many-body quantum fluctuations affecting data in the top panel have a visible effect in the transient relaxation dynamics: in fact, while the \emph{single-body} collective spin oscillations in the bottom panel become increasingly long-lived in larger systems as the classical limit is approached, the \emph{many-body} spatially-decaying interactions in the top panel cause damping and, correspondingly, are expected to lead to a Gibbs-type Floquet pre-thermal state in terms of the approximate high-frequency Floquet Hamiltonian (given by Eq. \eqref{eq:effXY_alpha} with the couplings $J/|i-j|^\alpha$ replaced by the experimental ones $J_{ij}$) \cite{prethermhighfreq}.  

\smallskip

\section{Conclusions and perspectives} 

Dynamical stabilization 
is a well understood mechanism since the original work  of P. Kapitza in 1951 \cite{Kapitza} on the driven classical pendulum, and has found several experimental applications, e.g. in particle synchrotrons \cite{synchrotron} and in Paul traps \cite{Paul}.  
More recently, it has been applied to quantum many-body physics out of equilibrium, for instance in order to stabilize Bose-Einstein condensate clouds in two and three spatial dimensions \cite{stabilizedBEC}, to prevent spin mixing in spinor condensates \cite{spinmixing,Kapitzaexp}, and to stabilize $\pi$-modes in driven bosonic Josephson junctions \cite{JosephsonKapitza}.
In all these cases, however, the dynamical stabilization by periodic driving involves a \emph{single} collective mode of the system. The driven infinite-range model which we preliminarily study in Sec. \ref{sec:infiniterange} represents a new instance of the same class.

On the other hand, the 
present work demonstrates that this dynamical stabilization can  occur in the presence of a rather general class of  multi-particle interactions, 
for which the many-body problem cannot be reduced to that of a single collective degree of freedom. In this respect, the mechanism investigated here is a bona-fide realization of the many-body Kapitza pendulum. Moreover, we have shown that \emph{ordered} ferromagnetic phases can also be stabilized via periodic driving, in a way that has no counterpart in equilibrium.

We focused here on the physics of long-range interacting spin chains to make a contact with current experiments with ion traps \cite{Monroe,Blatt}, which can be prepared in fully polarized initial states and whose time-evolution can be modeled by Hamiltonians of the form \eqref{eq:generalIsing}. 
The phenomena that we report are actually even more stable in higher-dimensional systems \cite{Browaeys,ions2d} and/or with higher spins \cite{ionsspin1}, where fluctuations are less effective.
In this respect, we emphasize that the dynamically stabilized non-equilibrium phases studied in this work can be observed even in relatively small systems with size $N\gtrsim 16$, accessible to exact numerical simulations  and well within the reach of state-of-the-art experiments, as demonstrated in Fig. \ref{fig:experimental}.

 We expect that our analysis can be generalized to other important phase transitions involving more complex symmetries, such as in the case of superconductors, providing access to a number of novel non-equilibrium phases of matter. In addition, our methodology can be extended in order to account for the effects of disorder and of external environments, further corroborating the robustness of our findings. 

 \section*{ Acknowledgments} 
 We thank M. Eckstein, G. Pagano and A. Polkovnikov for interesting discussions.
 We thank the authors of Ref.~\cite{Monroe} for kindly providing us with the data characterizing the interactions in their experimental setup.
J. M. is supported by the European Union's Horizon 2020 research and innovation programme under the Marie Sklodowska-Curie grant agreement No 745608 (QUAKE4PRELIMAT).

\vspace{0.2cm}

\appendix

\section{ Effective (Floquet) Hamiltonian} 
\label{app:Magnus}
 
Whenever the time-dependent Hamiltonian of a system has a 
period $T$, i.e., $H(t+T)=H(t)$, the resulting time-evolution operator $U(t_2,t_1)$ satisfies
%
\beq
U(t_0+nT,t_0) = 
\left[ U(t_0+T,t_0) \right]^n
\eeq
 for any integer $n$. Accordingly, it is 
 convenient to define an effective static Hamiltonian $H_{\text{eff}}$ \cite{SpagnoliReview,BukovReview},
\beq
U(t_0+T,t_0) = \mathcal{T} e^{ 
-i\int_{t_0}^{t_0+T} d\tau \, H(\tau) 
} \equiv e^ {-i T H_{\text{eff}}},
\eeq
usually referred to as the \textit{Floquet Hamiltonian}.
Its spectrum is defined up to integer multiples of the frequency $2\pi/T$ and it is 
independent of the choice of the reference time $t_0$. 
The state of the system at stroboscopic times $t_n=t_0+nT$ is therefore entirely and unambiguously determined by the  Floquet Hamiltonian $H_{\text{eff}}$.
A series expansion of $H_{\text{eff}}$ in powers of the period $T$,  known as the \textit{Magnus expansion}, 
can be written as 
\beq
\label{eq:Magnus}
H_{\text{eff}} = \sum_{n=0}^{\infty} H_{\text{eff}}^{(n)},
\eeq
with $H_{\text{eff}}^{(n)}$ proportional to $T^n$. Explicitly, the first terms read
\begin{align}
\label{eq:zerothorder}
H_{\text{eff}}^{(0)} &= \int_{t_0}^{t_0+T} \frac{d\tau_1}{T} \, H(\tau_1) ,\\ 
H_{\text{eff}}^{(1)} &= \frac{T}{2}  \int_{t_0}^{t_0+T}  \frac{d\tau_1}{T} \int_{t_0}^{t_0+\tau_1}  \frac{d\tau_2}{T} \, \big[ H(\tau_1),H(\tau_2)\big] ,
\label{eq:firstorder}
\end{align}
with the higher order terms involving a increasing number of nested commutators of $H$ at different times.
This expansion is convergent when $T$ is smaller than the inverse maximal extension of the spectrum of $H(t)$ \cite{SpagnoliReview}. 
 
%
We consider in the following 
the general class of systems defined in Eq.~\eqref{eq:generalIsing}, which encompasses 
the long- and infinite-range Ising chains 
subject to the effect of the periodic driving in Eq.~\eqref{eq:protocol}. 
In the simplest high-frequency limit $\Omega\to\infty$, the effective evolution is 
governed by the time-averaged Hamiltonian [cf. Eq.~\eqref{eq:zerothorder}], since the system has no time to react to variations of the external parameters much faster than its characteristic dynamical time scales. Nevertheless, if the modulation amplitude $\delta B$ is simultaneously increased with fixed 
$\zeta \equiv \delta B/\Omega$, 
the effective dynamics becomes qualitatively different from the former. Such qualitative changes involve a resummation of 
the high-frequency expansion \eqref{eq:Magnus} of the Floquet Hamiltonian \cite{BukovReview}. 
In some cases, an analytic solution in closed form can be obtained by performing a convenient time-periodic change of coordinates \cite{BukovReview}. Indeed, by moving into the oscillating frame 
\beq
\begin{pmatrix}
\sigma^x_{i} \\
\sigma^y_{i} \\
\sigma^z_{i} 
\end{pmatrix}=\begin{pmatrix} \cos\big( 2\zeta \sin(\Omega t) \big) \sigma'^x_{i} + \sin\big( 2\zeta \sin(\Omega t) \big) \sigma'^y_{i} \\
-\sin\big( 2\zeta \sin(\Omega t) \big) \sigma'^x_{i} + \cos\big( 2\zeta \sin(\Omega t) \big) \sigma'^y_{i} \\
 \sigma'^z_{i}
 \end{pmatrix},
\eeq
the time-periodicity of the external magnetic field is eliminated, and the proper equations of motion for $\vec{\sigma}'$ 
are generated by $\widetilde{H}(t) $, the static part of the Hamiltonian alone, which takes the same form as Eq.~\eqref{eq:generalIsing} with $\sigma^x_{i}\sigma^x_{j}$ replaced by
\beq
\begin{split}
&\cos^2\big( 2\zeta\sin(\Omega t) \big) \sigma'^x_{i} \sigma'^x_{j} + \sin^2\big( 2\zeta \sin(\Omega t) \big) \sigma'^y_{i} \sigma'^y_{j}  \\ 
& + \cos\big( 2\zeta \sin(\Omega t) \big) \sin\big( 2\zeta \sin(\Omega t) \big) \big( \sigma'^x_{i} \sigma'^y_{j} + \sigma'^y_{i} \sigma'^x_{j}\big)
\end{split}
\eeq
%
%
Crucially, the modulation $\delta B$ 
now intervenes only via the 
finite combination $\zeta$.  
A standard high-frequency expansion for the new time-periodic Hamiltonian $\widetilde{H}(t)$ will then reproduce the correct  high-frequency effective Hamiltonian $H_{\text{eff}}$. 
To lowest order, time-averaging yields the $XY$-model of Eq. \eqref{eq:effXY_alpha}
%
%
with an engineered anisotropy parameter $\gamma=\gamma(\zeta)=\mathcal{J}_0(4\zeta)$,
where $\mathcal{J}_0$ is the standard Bessel function of the first kind. 
%
%

\section{Finite-frequency corrections 
} 
\label{app:FiniteFreq}

We discuss here the correction of order $\Omega^{-1}$ to the effective Hamiltonian \eqref{eq:effXY_0}  within the Magnus expansion in the oscillating frame (see Appendix \ref{app:Magnus}). By using Eq. \eqref{eq:firstorder}, a straightforward calculation yields
\beq
\begin{split}
\mathcal{H}_{\text{eff}}^{(1)} = & \frac{8\pi \tilde{J}_0^2}{\Omega}\bigg[
  \kappa_{x^2,y^2}(\zeta) \, S_x S_y S_z -  \kappa_{xy,x^2}(\zeta) \, S_x^2 S_z  \\ &+  \kappa_{xy,y^2}(\zeta) \, S_y^2 S_z + \frac{B_0}{2\tilde{J}_0} \kappa_{z,xy}(\zeta) \, (S_y^2 - S_x^2)
\bigg],
\label{eq:effXY_1}
\end{split}
\eeq
in terms of the dimensionless coefficients $\kappa$
\beq
\begin{split}
\kappa_{x^2,y^2}(\zeta) & = \int_0^{2\pi} \frac{d\xi}{2\pi} \, ( \dot{A} B - \dot{B} A ), \\
\kappa_{xy,x^2}(\zeta) & = \int_0^{2\pi} \frac{d\xi}{2\pi} \, ( \dot{C} A - \dot{A} C ), \\
\kappa_{xy,y^2}(\zeta) & = \int_0^{2\pi} \frac{d\xi}{2\pi} \, ( \dot{C} B - \dot{B} C ), \\
\kappa_{z,xy}(\zeta) & = \int_0^{2\pi} \frac{d\xi}{2\pi} \, C, \\
\end{split}
\eeq
where the dots stand for derivatives with respect to the argument $\xi$, and
\beq
\begin{split}
A(\xi) & \equiv  \int_0^{\xi} \frac{d\eta}{2\pi} \, \cos^2(2\zeta \sin\eta), \\
B(\xi) & \equiv  \int_0^{\xi} \frac{d\eta}{2\pi} \, \sin^2(2\zeta \sin\eta) = \frac{\xi}{2\pi} - A(\xi) \\
C(\xi) & \equiv  \int_0^{\xi} \frac{d\eta}{2\pi} \, \cos(2\zeta \sin\eta) \sin(2\zeta \sin\eta). \\
\end{split}
\eeq
The classical Hamiltonian $\mathcal{H}_{\text{eff}}^{(0)}+\mathcal{H}_{\text{eff}}^{(1)}$, with $\mathcal{H}_{\text{eff}}^{(0)}$ given by Eq. \eqref{eq:effXY_0} and $\mathcal{H}_{\text{eff}}^{(1)}$ by Eq. \eqref{eq:effXY_1}, defines phase space trajectories on the Bloch sphere which better approximate the stroboscopic evolution of the collective spin at finite driving frequency $\Omega$ compared to the high-frequency limit \eqref{eq:effXY_0}. In particular, it is possible to determine the shift of the boundaries in Fig. \ref{fig:2} to order $\Omega^{-1}$.
We note that although the Magnus expansion is divergent \cite{BukovReview}, its truncations provide an increasingly accurate approximation of the regular orbits (KAM tori) in phase space, and hence of the dynamically stabilized phases.

%
\section{ Time-dependent spin wave theory} 
\label{app:TDSWT}
We briefly outline here this method, developed in Ref.  \cite{LeroseShort}, which provides a natural  and viable approach to investigate the equilibrium phases and the non-equilibrium evolution of a wide class of spin models. 
%
A time-dependent reference frame $\mathcal{R}=(\hat{X},\hat{Y},\hat{Z})$ is introduced, with its $\hat{Z}$-axis following the collective motion of $\vec{S}(t)$. 
The change of frame is implemented by a time-dependent global rotation operator 
parameterized by the spherical angles $\theta(t)$ and $\phi(t)$, whose evolution will be self-consistently determined in such a way that $S_X(t)\equiv S_Y(t)\equiv 0$.
For $\alpha=0$, when $H$ is a function of the total spin $\vec{S}$ only,
this requirement translates into a closed pair of ordinary differential equations for the two angles, as in Fig. \ref{fig:1}.
For $\alpha > 0$, the dependence of the interactions on the distance renders $H$ a function of not only the total spin, i.e., the $k=0$ Fourier mode of the spins, but also of all the $k$-modes of the spins, which now contribute to the dynamics.
In order to systematically take into account these fluctuations, the $\vec{\sigma}_i$ spins' deviations from the instantaneous direction of the $\hat{Z}$-axis are mapped to bosonic  variables $q_i,p_i$ via Holstein--Primakoff transformations. 
The out-of-equilibrium dynamics of the system governed by the Hamiltonian~\eqref{eq:generalIsing} with $\alpha\ne0$ involves quantum corrections to the classical evolution of the total spin $\vec{S}(t)$, which are expressed in terms of the corresponding spin wave variables  $\tilde{q}_k$, $\tilde{p}_k$. Retaining up to quadratic terms in the expansion (i.e., neglecting collisions among spin waves), one finds the evolution equations
\begin{equation}
\label{eq:vacuummotiondelta}
\left\{
\begin{split}
\frac{d\theta}{dt} =&\, 4 \left[ \tilde{J}_0   (1-\epsilon(t)) -  \delta^{pp}(t) \right]  \sin\theta \cos\phi \sin\phi  \\
                         & \quad\quad \quad +4 \, \delta^{qp}(t) \, \cos\theta \sin\theta \cos^2\phi ,\\
\frac{d\phi}{dt}  =& -2 B(t)  + 4 \left[\tilde{J}_0 (1-\epsilon(t)) -  \delta^{qq} (t)\right] \cos\theta \cos^2\phi  \\
                                          &\quad\quad\quad  +4  \, \delta^{qp} (t)\, \sin\phi \cos\phi   ,
\end{split}
\right.
\end{equation}
where $\delta^{\alpha\beta}(t) \equiv  \int_{-\pi}^\pi  \frac{dk}{\pi} \, \tilde{J}_k \Delta^{\alpha\beta}_k $ with $\alpha, \beta \in \{p,q\}$ is the quantum ``feedback'' in terms of correlation functions of the spin waves,
$
\Delta^{qq}_k (t) =  \left\langle \tilde{q}_k(t) \tilde{q}_{-k}(t) \right\rangle
$
and analogously $\Delta^{qp}_k$, $\Delta^{pp}_k$. The evolution of $\Delta^{\alpha\beta}_k$ is ruled 
by a  system of differential equations involving  $\theta(t)$ and $\phi(t)$ \cite{LeroseShort, LeroseLong}.
The validity of the quadratic approximation is controlled by the density of spin waves,
$
\epsilon(t)\equiv \int_{-\pi}^\pi  \frac{dk}{2\pi} \, (\Delta^{qq}_k + \Delta^{pp}_k -1) 
$
(with abuse of notation, here and in the main text we omit the brackets in denoting the quantum expectation values $\langle \vec{S}(t) \rangle$ and $\langle \epsilon(t) \rangle$).
The length of the collective spin $\lvert \vec{S}(t) \rvert = 1- \epsilon(t)$ is conserved by the dynamics only when $\alpha=0$. The approximation is justified as long as the density of excited spin waves 
is low, i.e., $\epsilon(t)\ll1$.
Initial fully polarized, spin-coherent states, as considered in Fig. \ref{fig:3}, correspond to
the initial data for Eqs.~\eqref{eq:vacuummotiondelta} 
$\theta(0) =  \theta_0$, $\phi(0) =\phi_0$ with $\Delta^{qq}_k(0) = \Delta^{pp}_k(0) = 1/2$, and $\Delta^{qp}_k(0) = 0$. In particular, $\epsilon(0)=0$.

\section{Coexistence of Kapitza and ferromagnetic phases} 
\label{app:CoexistenceKF}

\begin{figure}[t]
\centering
\includegraphics[width=0.46\textwidth]{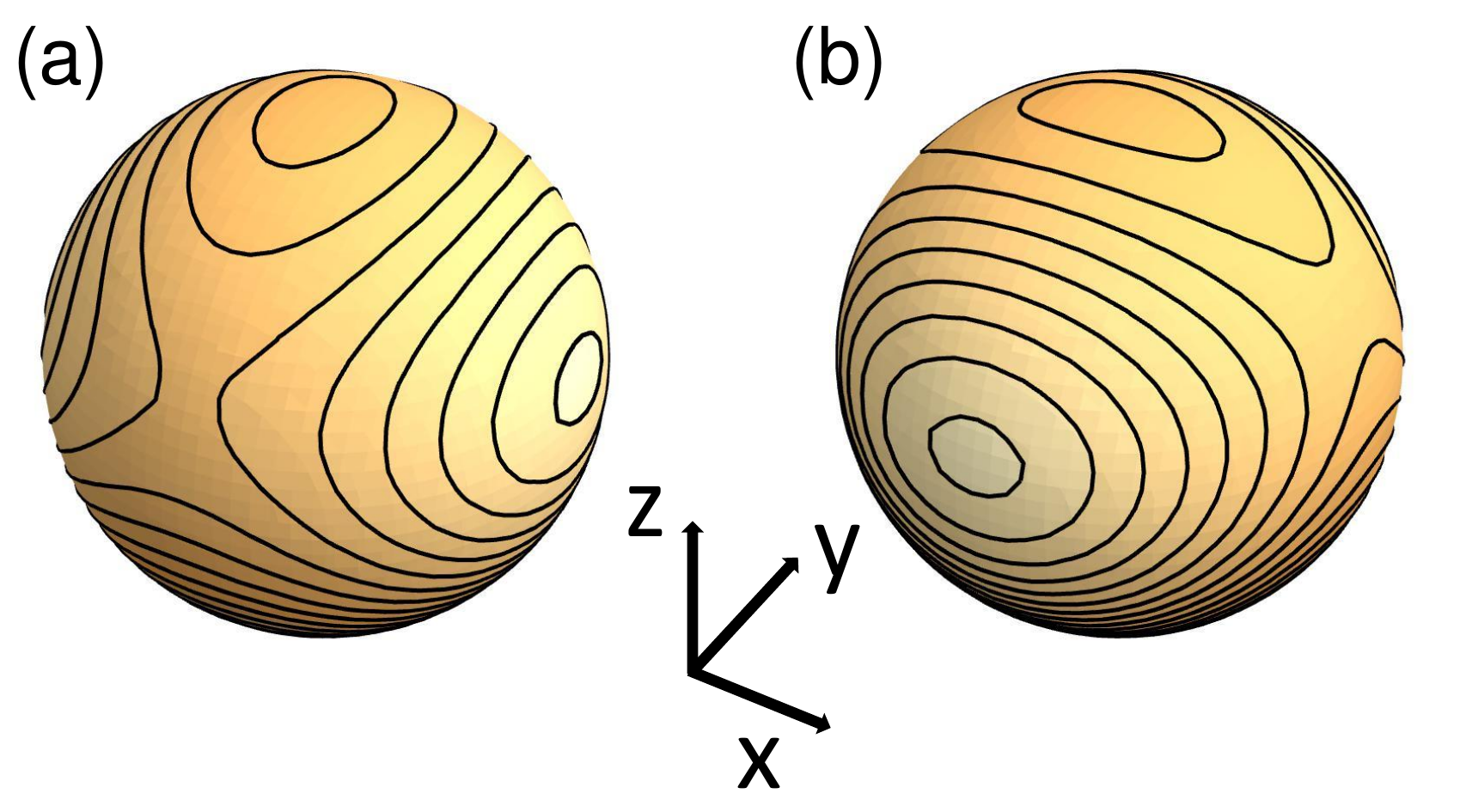}
\caption{%
Schematic phase portraits of the effective Hamiltonian $\mathcal{H}_{\text{eff}}$ in Eq.  \eqref{eq:effXY_0} on the sphere, with parameters belonging to the shaded region of the non-equilibrium phase diagram in Fig. \ref{fig:2}, 
corresponding to the coexistence of a dynamically stabilized Kapitza phase and the ferromagnetic phase $F_{\parallel}$ [(a), shaded blue in Fig. \ref{fig:2}], or $F_{\perp}$ [(b), shaded orange in Fig. \ref{fig:2}]. We emphasize that the paramagnetic configuration is here associated with a \emph{maximum} of $\mathcal{H}_{\text{eff}}$.
}
\label{fig:coexistence}
\end{figure}

\begin{figure}[t]
\centering
\includegraphics[width=0.46\textwidth]{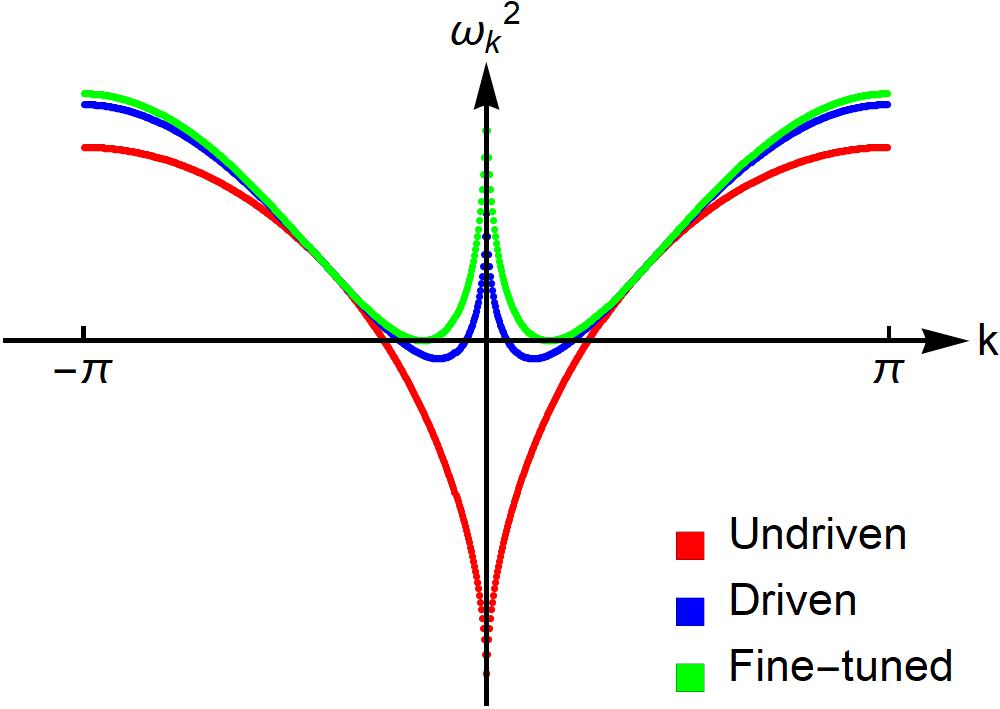} 
\caption{%
Effective spectrum of the quantum spin wave excitations 
around the unstable paramagnetic configuration for $\alpha=1.5$, $N=400$, $B_0/\tilde{J}_0=0.35$, in the presence of a high-frequency drive with $\delta B/\Omega=0$  (red), $0.4023$ (blue) and $0.6014$ (green), corresponding  to effective anisotropy parameters $\gamma=1$, $0.45$, and $0$, respectively. The blue and green points correspond to parameters within the shaded region in Fig. \ref{fig:2}, in which coexistence of Kapitza and ferromagnetic phases occurs in the infinite-range model. Although the collective $k=0$ mode is dynamically stabilized, for $\alpha\ne0$  an extended interval in the Brillouin zone appears with modes characterized by imaginary frequencies $\omega_k^2 <0$, as shown, e.g., by the blue points. This instability disappears only at isotropic points $\zeta_1,\zeta_2,\dots$ for which $\gamma=0$ [corresponding to the zeros of the Bessel function, see after Eq. \eqref{eq:effXY_0}], as shown by the green points.
}
\label{fig:coexistenceband}
\end{figure}

We briefly discuss here the coexistence of a second dynamically stabilized phase $K$ and the ferromagnetic phases $F_{\parallel,\perp}$ in the periodically driven infinite-range Ising model (cf. Fig. \ref{fig:2}), and the instability of the former for $\alpha\ne0$.

For $B_0 < \tilde{J}_0 (1-\lvert\gamma(\zeta)\rvert)$ (i.e., within the shaded region in Fig. \ref{fig:2}), the effective Hamiltonian \eqref{eq:effXY_0} has a \emph{maximum} at the paramagnetic point in addition to the two ferromagnetic minima in the $xz$- or $yz$-plane, depending on $\gamma$ being positive or negative, respectively. The corresponding phase portraits are shown in Fig. \ref{fig:coexistence}. 
Stable trajectories exist around the direction of both the ferromagnetic minima and the paramagnetic configuration, which would be unstable in the absence of the driving. 

Even though dynamical stabilization of the paramagnetic configurations occurs for the collective $k=0$ mode within this region of the parameters, a subset of the remaining modes with finite wavelengths $k\ne0$ turn out to be unstable when $\alpha>0$. In fact, within a linear stability analysis, the effective spectrum of excitations is given by
\beq
\omega_k^2 = 4 \big\{B_0-[1-\gamma(\zeta)] \, \tilde{J}_k\big\}  \big\{ B_0-[1+\gamma(\zeta)] \, \tilde{J}_k\big\},
\eeq 
as obtained by expanding Eq. \eqref{eq:effXY_alpha} in spin wave operators around $\theta=0$, and it features an interval of $k$ in the Brillouin zone with imaginary frequencies within the range of parameter values $B_0 < \tilde{J}_0 (1-\lvert\gamma(\zeta)\rvert)$ under consideration, see Fig. \ref{fig:coexistenceband}. The amplitude of this interval shrinks to zero when the anisotropy $\gamma=\mathcal{J}_0(4\zeta)$ approaches $0$, i.e., when the driving strength $\zeta$ equals one of the zeros $\zeta_n$ with $n=1,2,\dots$ of the Bessel function. Away from this discrete set of values, the Kapitza phase coexisting with the ferromagnetic phases turns out to be destabilized by these finite-wavelength fluctuations, at least at the level of linear stability, although the collective $k=0$ mode is 
dynamically stabilized.

It is interesting to note that when $\zeta$ is tuned to an isotropic point $\zeta_n$, the many-body Kapitza phase discussed above becomes stable in the high-frequency limit $\Omega\to\infty$, and thus approximately stable for $ \Omega \gg \tilde{J}_0 $. The reason behind such stability can be easily traced back to the stroboscopic conservation of $S_z$: indeed, if the system is initialized in a fully polarized state with a small displacement $\theta_0$ from the $z$ axis, the collective spin has to remain trapped in a neighborhood of the otherwise unstable configuration $\theta=0$, because $S_z(t_n=nT) \approx 1- \theta_0^2/2$ cannot decrease.

\section{Non-equilibrium phase diagram for $\alpha>0$} 
\label{app:shift}

Here we discuss the modification of the phase boundaries in Fig. \ref{fig:2} when $\alpha$ increases from $0$ to $2$ and beyond, leading to Fig. \ref{fig:NEQphasediagramAlpha}. These boundaries are determined by the critical value $B_0=B_{\text{cr}}(\zeta)$ below which the high-frequency effective Hamiltonian \eqref{eq:effXY_alpha} develops ferromagnetic ordering, either in the $xz$- or in the $yz$-plane depending on the anisotropy $\gamma(\zeta)$ being positive or negative, respectively. 
Ferromagnetic ordering arises as soon as tilted spin configurations, with average orientation forming an angle $\theta\ne0$ with the field direction $z$, acquire a lower energy than the paramagnetic states with $\theta=0$.

In the infinite-range case $\alpha=0$, the critical line is determined by $B_{\text{cr}}(\zeta) = \tilde{J}_0(1+\lvert \gamma(\zeta) \rvert)$, in correspondence of which the energy landscape $\mathcal{E}_{\text{eff}}(\theta) \equiv \langle H_{\text{eff}}\rangle_{\theta,\phi^*}$ (with $\phi^*=0$ or $\pi/2$ depending on $\gamma(\zeta)>0$ or $<0$, respectively) of the semiclassical collective spin $\vec{S} = (\sin\theta \cos\phi, \sin\theta\sin\phi,\cos\theta)$ changes from a single well for $B_0 > B_{\text{cr}}(\zeta) $, with the minimum at $\theta^*=0$, to a symmetric double well for $B_0 < B_{\text{cr}}(\zeta) $, with the minima at $\theta^*= \pm  \arccos (B_0 / B_{\text{cr}}(\zeta)) $.

In the presence of an interaction with $\alpha\ne0$, the quantum fluctuations of all the spin degrees of freedom with Fourier wavevector $k\ne0$ modify the effective energy landscape $\mathcal{E}_{\text{eff}}(\theta)$ of the collective spin, thereby shifting the critical value of $B_0$ to $B_{\text{cr}}(\zeta) \equiv \tilde{J}_0(1+\lvert \gamma(\zeta) \rvert) + \Delta B_{\text{cr}}(\zeta)$. Within the spin wave treatment introduced in Refs. \cite{LeroseShort,LeroseLong}, we can compute $\mathcal{E}_{\text{eff}}(\theta)$ and hence the shift $\Delta B_{\text{cr}}(\zeta)$ to lowest order in $\tilde{J}_{k\ne0}$, via a variational approach. 
In particular,
we consider the expansion of $H_{\text{eff}}$ to quadratic order in the spin wave operators around the direction in the $xz$- or $yz$-plane identified by the angle $\theta$, resulting in Eq. \eqref{eq:timeindepH} with $\phi=0$ or $\pi/2$ respectively, and we determine the energy landscape $\mathcal{E}_{\text{eff}}(\theta) \equiv \min_{\psi} \langle H_{\text{eff}}\rangle_{\psi}$ by computing the parametric spin wave ground state energy. We obtain
\beq
\label{eq:energylandscape}
\frac{\mathcal{E}_{\text{eff}}(\theta)}{N} = -\tilde{J}_0\frac{1+\lvert\gamma\rvert}{2} \sin^2\theta - B_0 \cos\theta  
+ \int_{-\pi}^\pi  \frac{dk}{\pi} \, \frac{\omega_k-\omega_k^{(0)}}{2},
\eeq
where
\beq
\begin{split}
\omega_k^2 = & 4 \Big[ \tilde{J}_0  (1+\lvert\gamma\rvert) \sin^2\theta - \tilde{J}_k   (1+\lvert\gamma\rvert) \cos^2\theta + B_0 \cos\theta \Big] \\
& \qquad\quad \times \Big[ \tilde{J}_0  (1+\lvert\gamma\rvert) \sin^2\theta - \tilde{J}_k  (1-\lvert\gamma\rvert) + B_0 \cos\theta \Big]
\end{split}
\eeq
and 
\beq
\omega_k^{(0)} =  2 \Big[ \tilde{J}_0  (1+\lvert\gamma\rvert) \sin^2\theta + B_0 \cos\theta \Big] = \omega_k \Big\rvert_{\tilde{J}_{k\ne0}=0}.
\eeq
The last term in Eq. \eqref{eq:energylandscape} represents the zero-point contribution of quantum fluctuations with arbitrary $k$.
The energy landscape $\mathcal{E}_{\text{eff}}(\theta)$ in Eq. \eqref{eq:energylandscape} can be expanded at small $\theta$ as 
\beq
\mathcal{E}_{\text{eff}}(\theta) = \mathcal{E}_{\text{eff}}(\theta=0) + \omega_{\text{eff}}(B_0) \,  \frac{\theta^2}{2} +  \mathcal{O}\big(\theta^4\big),
\eeq 
and the critical value $B_{\text{cr}}$ of $B_0$ is determined by the equation $\omega_{\text{eff}}(B_0) = 0$, corresponding to the transition from a single ($\omega_{\text{eff}} > 0$) to a double ($\omega_{\text{eff}} < 0$) well landscape. The solution may be found by formally expanding $B_0$ in powers of $\tilde{J}_{k\ne0}$ and equating both sides order by order. This procedure yields
\beq
\begin{split}
B_{\text{cr}}(\zeta) = & \tilde{J}_0 \big(1+\lvert \gamma(\zeta) \rvert\big) \\
& \times \Bigg\{
1- 
\frac{2\lvert \gamma(\zeta) \rvert + 3 \lvert \gamma(\zeta) \rvert^2}{4 \big(1+\lvert \gamma(\zeta) \rvert\big)^2}
\Bigg[
\int_{-\pi}^\pi  \frac{dk}{\pi} \, \bigg( \frac{\tilde{J}_k}{\tilde{J}_0} \bigg)^2 
\Bigg]
\Bigg\} \\ &
+ \mathcal{O}\big(\tilde{J}_k^3 \big),
\end{split}
\eeq
which yields the negative quadratic correction $\Delta B_{\text{cr}}$ reported in Eq. \eqref{eq:shift}. This implies that the quantum fluctuations destabilize the ferromagnetic ordering, as expected on physical grounds. Moreover, we observe that the shift $\Delta B_{\text{cr}}$ of the critical value is maximal for $\gamma=1$ and vanishes at isotropic points with $\gamma=0$, as illustrated in Fig. \ref{fig:NEQphasediagramAlpha}. Note that this result is valid for any type of spatial dependence $J_{\lvert i-j \rvert}$ of the interactions in Eq. \eqref{eq:generalIsing}.

We finally remark that in the opposite limit $\alpha=\infty$, the system \eqref{eq:generalIsing} reduces to the standard quantum Ising chain with nearest-neighbor interactions (which has been studied in Refs. \cite{PDIsingchain,BrandesIsing}).
 In this case, the effective high-frequency Hamiltonian \eqref{eq:effXY_alpha} describes the XY quantum spin chain, which is exactly solvable in terms of free fermions \cite{Lieb}. Correspondingly, the quantum critical point $B_{\text{cr}}=\tilde{J}_0$ is independent of $\gamma$, and thus of the driving strength $\zeta$. Accordingly, it is natural to conjecture that the left boundary of the Kapitza phase in Fig. \ref{fig:NEQphasediagramAlpha} moves leftwards as $\alpha$ exceeds $1$, and eventually approaches the straight vertical line $B_{\text{cr}}(\zeta)=\tilde{J}_0$ when $\alpha\to\infty$. However, it is important to note that ferromagnetic ordering cannot arise at finite energy density for $\alpha>2$. Thus, in this case the equilibrium phase diagram of the effective Hamiltonian does not provide information on the dynamics in the presence of the driving. 


\vspace{0.2cm}



\vspace{0.2cm}


\end{document}